\documentclass[
superscriptaddress,
twocolumn,
floatfix,
amsmath,
amssymb,
aps,
prl,
]{revtex4-2}

\usepackage{bm}
\usepackage{graphicx}
\ifdefined\pdfsuppresswarningpagegroup
\fi

\newcommand{\vb}[1]{\bm{#1}}
\newcommand{\one}{\sigma_0}

\newcommand{\Tr}{\mathrm{Tr}}
\newcommand{\intk}{\int_{\vb{k}}\!}
\newcommand{\svec}{\vb{\sigma}}

\begin{document}

\title{Programmable Spin Conversion in Gradient Quantum Matter}

\author{Mamoru Matsuo}
\email{mamoru@ucas.ac.cn}
\affiliation{Kavli Institute for Theoretical Sciences, University of Chinese Academy of Sciences, Beijing 100190, China}
\affiliation{CAS Center for Excellence in Topological Quantum Computation, University of Chinese Academy of Sciences, Beijing 100190, China}
\affiliation{Advanced Science Research Center, Japan Atomic Energy Agency, Tokai 319-1195, Japan}
\affiliation{RIKEN Center for Emergent Matter Science (CEMS), Wako, Saitama 351-0198, Japan}

\author{Yuta Sekino}
\email{sekino.yuta.y2@f.mail.nagoya-u.ac.jp}
\affiliation{Institute for Advanced Research, Nagoya University, Nagoya 464-8601, Japan}
\affiliation{Department of Physics, Nagoya University, Furo-cho, Chikusa-ku, Nagoya, Aichi 464-8602, Japan}
\affiliation{Interdisciplinary Theoretical and Mathematical Sciences Program (iTHEMS), RIKEN, Wako, Saitama 351-0198, Japan}
\affiliation{Nonequilibrium Quantum Statistical Mechanics RIKEN Hakubi Research Team, RIKEN Cluster for Pioneering Research (CPR), Wako, Saitama 351-0198, Japan}

\author{Hiroyuki Tajima}
\email{hiroyuki.tajima@phys.s.u-tokyo.ac.jp}
\affiliation{Department of Physics, Graduate School of Science, The University of Tokyo, Tokyo 113-0033, Japan}
\affiliation{RIKEN Nishina Center, Wako 351-0198, Japan}
\affiliation{Quark Nuclear Science Institute, The University of Tokyo, Tokyo 113-0033, Japan}

\begin{abstract}
We propose programmable spin conversion in ultracold gases as gradient quantum matter, whose spin-dependent self-energy varies in space. Quantum kinetic theory shows that a dissipative self-energy curvature turns a force-driven scalar anisotropy into a spin source with mixed longitudinal-transverse momentum parity.
Spin-resolved time-of-flight imaging can reveal a transverse spin texture that changes sign when either the drive or programmed curvature is reversed.
Ultracold gases thereby offer a controllable spin source for gradient quantum matter.
\end{abstract}

\maketitle

\paragraph{Introduction---}
Understanding nonequilibrium spin transport is essential for revealing angular-momentum dynamics and controlling spin degrees of freedom.
In conventional solids, however, spin transport is often obscured by uncontrolled environments such as disorders and boundaries, making it difficult to isolate the intrinsic mechanisms governing nonequilibrium spin dynamics.

Ultracold gases provide a controlled setting for driving spin-dependent distributions.
A central established direction has been Hamiltonian engineering, including synthetic spin-orbit coupling (SOC) produced by Raman dressing and artificial gauge fields~\cite{Lin2011Science,Zhai2015Review}.
Spatially inhomogeneous spin control is now experimentally concrete: in these atomic systems, hyperfine states play the role of electron spin, enabling studies of spin transport in ultracold Fermi gases~\cite{PhysRevLett.101.150401,PhysRevLett.103.010401,sommer2011universal,sommer2011spin,PhysRevLett.109.050403,koschorreck2013universal,bardon2014transverse,krauser2014giant,Trotzky2015,krinner2016mapping,valtolina2017exploring,PhysRevLett.118.130405,PhysRevA.99.063620,PhysRevA.108.L041304,9ks8-zv9b,Royse2026}.
Meanwhile, optical potentials, near-resonant light, and dephasing fields impose spin-dependent coherent shifts with programmable spatial profiles~\cite{grimm2000optical,gauthier2016direct,PhysRevLett.115.073002,nichols2019spin}.
Spin-dependent dissipation has been realized experimentally~\cite{Corman2019PRA,ren2022chiral,zhao2025two}.
In particular, inhomogeneous spin-dependent dissipation has been implemented using a near-resonant optical tweezer in the context of a dissipative atomic point contact~\cite{Corman2019PRA}.
Along with the rapid progress of experimental techniques, even inhomogeneous interactions can be realized in ultracold Fermi gases~\cite{PhysRevLett.122.040405,PhysRevA.106.023322}.

Taken together, these experiments show that coherent and dissipative spin controls can now be implemented in ultracold gases.
Spatially inhomogeneous controllability has entered cold-atom spin physics mainly as the ability to prescribe local spin shifts and dissipations.
For spin transport, the same controllability can potentially be extended from local spin parameters to the nonequilibrium atomic distribution itself.
Accordingly, spin-gradient phenomena such as the Leggett-Rice effect~\cite{Trotzky2015} and chirality-induced spin selectivity~\cite{Royse2026} have been realized with artificial magnetic gradients.

\begin{figure}[!ht]
\centering
\includegraphics[width=0.97\columnwidth]{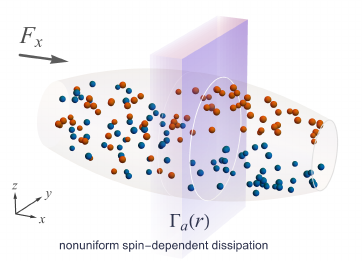}
\caption{Programmable gradient quantum spin material in a two-component cold-atom gas.
A spin-independent drive \(F_x\) carries atoms through a tube-shaped cloud crossed by a finite-thickness dissipative region with nonuniform spin-dependent \(\Gamma_a(\vb r)\).
The incoming components are mixed.
After crossing the region, the distribution is biased, with the \(+a\)-polarized atoms (orange) preferentially on the \(+z\) side and the \(-a\)-polarized atoms (blue) on the \(-z\) side.
This transverse spin imbalance illustrates scalar-to-spin conversion, the cold-atom analog of charge-to-spin conversion in an electronic gradient material.}
\label{fig:concept}
\end{figure}

Solid-state spintronics points to the same connection between spatial gradients and charge-to-spin conversion.
Efficient spin conversion has commonly been associated with strong band spin-orbit coupling in heavy-element spin Hall materials, Rashba interfaces, spin-orbit torques, and multilayers~\cite{Sinova2015,Manchon2019}.
Recent experiments show that sizable spin signals can arise even in weak-spin-orbit materials when spin-dependent material parameters vary in space, as reported for surface-oxidized metals, composition-gradient films, and spin-vorticity heterostructures~\cite{An2016NatCommun,Okano2019,Nakayama2023,Horaguchi2025,Yi2025PRL,Shi2026AFM}.
In such gradient materials, direct charge-to-spin and inverse spin-to-charge responses can differ by orders of magnitude~\cite{Okano2019,Horaguchi2025}.
Together, they suggest spin conversion in weak-spin-orbit gradient materials.
These observations connect to a broader weak-spin-orbit setting in which orbital angular-momentum dynamics underlies Hall transport, current-induced torque, and edge accumulation~\cite{Kontani2009PRL,GoJoKimLee2018PRL,JoGoLee2018PRB,GoLee2020PRR,Go2020PRR,ValetRaimondi2025PRB,Valet2025PRL}.
In solids, however, the relevant gradient is tied to interfaces, disorder, microscopic electronic structure, spin sinks, and electrical or spin-accumulation readout.
These make it difficult to distinguish which part of a measured signal is created by the local gradient source.

In this Letter, we demonstrate programmable gradient quantum matter for spin conversion in a two-component ultracold Fermi gas with coherent self-energy gradients and nonuniform spin-dependent dissipation.
Our central result is a dissipative-curvature spin source whose transverse momentum texture is parity-locked to the drive and to the programmed gradient, and is accessible to spin-resolved time-of-flight measurements with source-region selection.
In the quantum kinetic equation, the spatial self-energy profile enters the Moyal derivative: coherent first-gradient terms and dissipative curvature terms generate scalar-to-spin sources.
After momentum integration, the measured spin response is selected by the momentum weight associated with spin density, spin current, finite apertures, or momentum-resolved imaging.
The dissipative curvature source carries longitudinal-transverse parity information, allowing spin polarization and flow direction to be selected by the spin component, curvature direction, and readout.
Real-space density imaging and boundary or reservoir readout can access the spin-density profile and spin-current signals.
This separation keeps preparation, conversion, real-space density readout, and transport-current readout experimentally distinguishable.
In this way, ultracold atoms provide a programmable gradient quantum material for spin conversion based on spatial control of self-energies.

\paragraph{Two-Component Gas With Coherent and Dissipative Inhomogeneity---}
We consider a two-component Fermi gas with a field operator
\begin{equation}
\hat{\Psi}(\vb{r})=\binom{\hat{\psi}_{\uparrow}(\vb{r})}{\hat{\psi}_{\downarrow}(\vb{r})},\ \{\hat{\psi}_{\alpha}(\vb{r}),\hat{\psi}_{\beta}^{\dagger}(\vb{r}')\}=\delta_{\alpha\beta}\delta(\vb{r}-\vb{r}'),
\label{eq:field}
\end{equation}
where \(\alpha,\beta\) are spin indices.
Fermionic anticommutators are used for definiteness~\footnote{For a normal spinful bosonic gas, anticommutators of $\hat{\psi}_\alpha(\bm{r})$ are replaced by commutators and the equilibrium occupation reads the Bose distribution. The Moyal derivative structure generating the gradient and curvature sources is independent of quantum statistics. Response coefficients and any condensed component require the corresponding bosonic kinetic treatment.}.
The single-particle Hamiltonian is
\begin{align}
\hat H
&=
\int d\vb r\,
\hat\Psi^\dagger(\vb r)h(\vb r)\hat\Psi(\vb r),
\nonumber\\
h(\vb r)
&=
\left(
-\frac{\hbar^2\nabla^2}{2m}
-F_xx
+U_0(\vb r)\right)\one+\vb U(\vb r)\cdot\svec .
\label{eq:HS}
\end{align}
Here \(\sigma_0\) is the \(2\times2\) identity matrix, and \(\svec=(\sigma_x,\sigma_y,\sigma_z)\) denotes the vector of Pauli matrices.
The force \(F_x\) is spin independent and prepares the scalar nonequilibrium distribution; in what follows, scalar means spin independent.
The scalar and spin-dependent coherent potentials \(U_0(\bm{r})\) and \(\vb{U}(\vb{r})\) may be included in the Hermitian self-energy.
After engineered reservoirs, loss channels, or dephasing fields are eliminated, the retarded and advanced self-energies are written as
\begin{align}
\Sigma^{R,A}
&=
\Lambda_0\one+\vb\Lambda\cdot\svec
\mp\frac{i}{2}
\left(
\Gamma_0\one+\vb\Gamma\cdot\svec
\right).
\label{eq:selfenergy}
\end{align}
The vectors \(\vb\Lambda=(\Lambda_x,\Lambda_y,\Lambda_z)\) and \(\vb\Gamma=(\Gamma_x,\Gamma_y,\Gamma_z)\) denote the coherent spin-dependent shifts and spin-dependent dissipative rates, respectively.
For the self-energy, superscripts \(<\), \(>\), and \(K\) denote lesser, greater, and Keldysh components, with \(\Sigma^K=\Sigma^>+\Sigma^<\) in the convention used here.
In a local Markov representation, \(\Sigma^<=i\Gamma f^\Sigma\), \(\Sigma^>=-i\Gamma(\one-f^\Sigma)\), and \(\Sigma^K=-i\Gamma(\one-2f^\Sigma)\), with matrix multiplications in spin space understood.
Here \(f^\Sigma\) is the spin-matrix occupation supplied by the eliminated environment.
It is decomposed in the same scalar-spin form, \(f^\Sigma=f_0^\Sigma\one+\vb f^\Sigma\cdot\svec\).
The gas distribution and the environmental distribution need not coincide in a nonequilibrium state.

\paragraph{Spin Continuity From Self-Energy Gradients---}
We introduce
the greater and lesser Green functions \(G^>_{\alpha\beta}(1,2)=-(i/\hbar)\langle\hat\psi_\alpha(1)\hat\psi_\beta^\dagger(2)\rangle\) and \(G^<_{\alpha\beta}(1,2)=(i/\hbar)\langle\hat\psi_\beta^\dagger(2)\hat\psi_\alpha(1)\rangle\),
where \(1=(\vb r_1,t_1)\) and \(2=(\vb r_2,t_2)\).
The Keldysh component is \(G^K=G^>+G^<\), while \(G^R_{\alpha\beta}(1,2)=-(i/\hbar)\Theta(t_1-t_2)\langle\{\hat\psi_\alpha(1),\hat\psi_\beta^\dagger(2)\}\rangle\) and \(G^A_{\alpha\beta}(1,2)=(i/\hbar)\Theta(t_2-t_1)\langle\{\hat\psi_\alpha(1),\hat\psi_\beta^\dagger(2)\}\rangle\) are retarded and advanced components.
The Kadanoff-Baym equation for \(G^K\) is~\cite{KadanoffBaym1962,Keldysh1965,RammerSmith1986,HaugJauho2008}
\begin{equation}
(G_0^{-1}-\Sigma^{R})\circ G^{K}-G^{K}\circ(G_0^{-1}-\Sigma^{A})=\Sigma^{K}\circ G^{A}-G^{R}\circ\Sigma^{K},
\label{eq:KB}
\end{equation}
where \(G_0^{-1}\) is the inverse bare Green function, and \(\circ\) denotes convolution over intermediate space-time coordinates with spin-matrix multiplication.
We use center and relative variables \(X=(\vb r,t)\), \(\xi=(\vb r_-,t_-)\), \(\vb r=(\vb r_1+\vb r_2)/2\), \(\vb r_-=\vb r_1-\vb r_2\), \(t=(t_1+t_2)/2\), and \(t_-=t_1-t_2\).
With \(K=(\vb k,\omega)\) and \(\int_{\vb r_-,t_-}\equiv\int d\vb r_-dt_-\), the Wigner transform is
\begin{align}
A(X;K)
&=
\int_{\vb r_-,t_-}\!\!\!\! e^{-i\vb k\cdot\vb r_-+i\omega t_-}\!A(X+\xi/2;X-\xi/2).
\label{eq:wigner}
\end{align}
In the quasiparticle projection, the gas distribution is the \(2\times2\) spin matrix \(f(X;K)\) defined by \(G^<(X;K)=i\mathcal A_{\mathrm{sp}}(X;K)f(X;K)\) and \(G^>(X;K)=-i\mathcal A_{\mathrm{sp}}(X;K)[\one-f(X;K)]\), where \(\mathcal A_{\mathrm{sp}}=i(G^R-G^A)\) is the single-particle spectral function.
The Wigner transform converts the convolution into the Moyal product, with the Poisson bracket denoted by \(\{\cdots,\cdots\}_{\rm PB}\)
\begin{align}
A\circ B
&\rightarrow
A\star B
=
AB+\frac{i}{2}\{A,B\}_{\mathrm{PB}}+\cdots ,
\label{eq:moyal}\\
\{A,B\}_{\mathrm{PB}}
&=
\nabla_{\!r}A\cdot\nabla_{\!k}B
-\nabla_{\!k}A\cdot\nabla_{\!r}B
\nonumber\\
&\phantom{=}
-\partial_{\omega}A\,\partial_tB
+\partial_tA\,\partial_{\omega}B .
\label{eq:PB}
\end{align}
Unless stated otherwise, we set \(\hbar=1\) from here on.
After the quasiparticle projection and the \(\omega\) integration, the spin-matrix distribution \(f(\vb r,\vb k,t)\) obeys
\begin{equation}
\left(
\partial_t+\vb v_{\vb k}\cdot\nabla_{\vb r}
\right)f
+F_x\partial_{k_x}f
=\mathcal I[f],
\label{eq:qkt}
\end{equation}
where \(\vb{v}_{\vb{k}}\) is the group velocity and \(\mathcal I[f]\) is the collision integral given by
\begin{equation}
\mathcal I[f]=-i(\Lambda\star f-f\star\Lambda)-\frac{1}{2}(\Gamma\star\Delta f^\Sigma+\Delta f^\Sigma\star\Gamma),
\label{eq:collision}
\end{equation}
with \(\Delta f^\Sigma=f-f^\Sigma\), \(\Lambda=\Lambda_0\one+\vb\Lambda\cdot\svec\), and \(\Gamma=\Gamma_0\one+\vb\Gamma\cdot\svec\).
Here \(f^\Sigma\) is the distribution supplied by eliminated reservoirs, loss channels, or dephasing fields,
and \(\Delta f^\Sigma=f-f^\Sigma\) measures its mismatch with the gas distribution.
We decompose the distribution as
\begin{align}
f(\vb r,\vb k,t)
&=
f_0(\vb r,\vb k,t)\one
+\vb f(\vb r,\vb k,t)\cdot\svec ,
\nonumber\\
f_0
&=
f_{\rm eq}(\varepsilon_{\vb k})
+\delta f_0(\vb r,\vb k,t)
,
\label{eq:fdecomp}
\end{align}
where \(f_0\) and \(\vb f=(f_x,f_y,f_z)\) are scalar and spin components, respectively, and \(\varepsilon_{\vb k}\) is the kinetic energy.
For an unpolarized reference environment with \(f^\Sigma=f_{\rm eq}\one\), the nonequilibrium part of the mismatch is
\(\delta(\Delta f^\Sigma)=\delta f_0\,\one+\vb f\cdot\svec\).
Thus \(\vb f\) denotes the nonequilibrium spin distribution, whereas only the scalar component is split into \(f_{\rm eq}+\delta f_0\).

We expand Eq.~\eqref{eq:collision} through second order in spatial gradients.
For static, momentum-independent self-energies, the coherent and dissipative contributions of $\mathcal{I}[f]$ are given by
\begin{align}
\mathcal I_\Lambda
&=
-i[\Lambda,f]
+\frac{1}{2}\{\partial_i\Lambda,\partial_{k_i}f\}
\nonumber\\
&{}
+\frac{i}{8}[\partial_i\partial_j\Lambda,\partial_{k_i}\partial_{k_j}f]
+O(\nabla_r^3),
\nonumber\\
\mathcal I_\Gamma
&=
-\frac{1}{2}\{\Gamma,\Delta f^\Sigma\}
-\frac{i}{4}[\partial_i\Gamma,\partial_{k_i}\Delta f^\Sigma]
\nonumber\\
&{}
+\frac{1}{16}
\{\partial_i\partial_j\Gamma,
\partial_{k_i}\partial_{k_j}\Delta f^\Sigma\}
+O(\nabla_r^3).
\label{eq:expandedCollision}
\end{align}
The coherent first-gradient term is an anticommutator and can convert a scalar distribution into a spin response.
The dissipative first-gradient term is a spin-space commutator and therefore does not generate spin from a scalar input in a local Markov model.

To analyze scalar-to-spin conversion, we project Eq.~\eqref{eq:qkt}
onto its spin components while retaining \(f_0\) as the scalar driving distribution.
For the spin-projected equations below, \(a,b,c\in\{x,y,z\}\) label Cartesian spin components, and repeated spin indices are summed.
Taking the spin-\(a\) trace ${\rm Tr}\{\sigma_a\cdots\}$ of Eq.~\eqref{eq:qkt} gives
\begin{align}
2&\left(\partial_t+\vb v_{\vb k}\cdot\nabla_{\vb r}\right)f_a
+2F_x\partial_{k_x}f_a
\nonumber\\
&=
\Tr\{\sigma_a\mathcal I_{\mathrm{loc}}[f]\}
+2\mathcal K_a^{\Lambda(1)}
+2\mathcal K_a^{\Gamma(1)}
+2\mathcal K_a^{\Gamma(2)}
+O(\nabla_{\vb r}^3),
\label{eq:spinProjected}
\end{align}
where \(\mathcal I_{\mathrm{loc}}=-i[\Lambda,f]-\{\Gamma,\Delta f^\Sigma\}/2\), and
\begin{align}
\mathcal K_a^{\Lambda(1)}
&=
\left(\partial_i\Lambda_0\right)\partial_{k_i}f_a
+\left(\partial_i\Lambda_a\right)\partial_{k_i}f_0,
\label{eq:kernelLambda}\\
\mathcal K_a^{\Gamma(1)}
&=
\frac{1}{2}\epsilon_{abc}
\left(\partial_i\Gamma_b\right)
\partial_{k_i}\Delta f_c^\Sigma,
\label{eq:kernelGamma}\\
\mathcal K_a^{\Gamma(2)}
&=
\frac{1}{8}
\left[
\left(\partial_i\partial_j\Gamma_a\right)
\partial_{k_i}\partial_{k_j}\Delta f_0^\Sigma
+
\left(\partial_i\partial_j\Gamma_0\right)
\partial_{k_i}\partial_{k_j}\Delta f_a^\Sigma
\right].
\label{eq:kernelGammaSecond}
\end{align}
For the scalar input used below, the coherent second-gradient term in Eq.~\eqref{eq:expandedCollision} vanishes pointwise because
\([\partial_i\partial_j\Lambda,\partial_{k_i}\partial_{k_j}f_0\one]=0\).
For a spin component generated at first order in the programmed gradients, this term is \(O(\nabla_{\vb r}^3)\) and lies beyond the present truncation; for an independently prepared zeroth-order spin polarization, however, its spin projection is \(O(\nabla_{\vb r}^2)\) and must be restored together with the second term of Eq.~\eqref{eq:kernelGammaSecond}.
For scalar input, \(\Delta f^\Sigma=\Delta f_0^\Sigma\one\), Eq.~\eqref{eq:kernelGamma} vanishes, whereas the first term of Eq.~\eqref{eq:kernelGammaSecond} is the leading gradient-specific dissipative scalar-to-spin curvature source.
The second term in Eq.~\eqref{eq:kernelGammaSecond} is retained for completeness and describes a scalar dissipative curvature acting on an already spin-polarized distribution.
We keep the spin component of the force drift explicitly in Eq.~\eqref{eq:spinProjected}.
For full spin-density integration in the clean continuum, \(\intk F_x\partial_{k_x}f_a\) is a \(k\)-space boundary term and vanishes when the distribution decays sufficiently fast or Brillouin-zone boundary contributions cancel.
In the dissipative response below, the same spin-independent force is retained through the scalar nonequilibrium correction \(\delta f_0^{(F_x)}\).
Momentum integration of Eq.~\eqref{eq:spinProjected} gives
\begin{align}
\partial_t s_a+\partial_i j_{s,i}^{a}
&=
\mathcal S_a^{\mathrm{loc}}
+\mathcal S_a^{\Lambda(1)}
+\mathcal S_a^{\Gamma(1)}
+\mathcal S_a^{\Gamma(2)},
\label{eq:spinContinuity}
\end{align}
where \(s_a=\intk f_a\), \(j_{s,i}^{a}=\intk v_if_a\), \(\mathcal S_a^{X}=\intk\mathcal K_a^X\), and \(\mathcal S_a^{\mathrm{loc}}\) contains local coherent torque, local spin-selective loss or gain, and phenomenological spin relaxation, respectively (for convenience, we omitted $\bm{k}$ in $v_{\bm{k},i}$ as $v_i$).
Equation~\eqref{eq:spinContinuity} is written after momentum integration: whether a gradient source survives a particular measurement is decided by the momentum weight selected by the observable, as shown in the End Matter.
This balance plays a preparatory role, since the spin-dipole evolution follows from it; the momentum-resolved kernels remain the primary objects and are analyzed next.

\paragraph{Reading Out the Momentum-Space Source---}
Here we specify the primary observable of this Letter, the momentum-resolved spin texture, and organize the integrated density, current, and higher moments as momentum-weighted projections of the source kernels.
For many-body clouds, ballistic time-of-flight imaging maps the initial momentum distribution to the expanded density in the far-field limit~\cite{Bloch2008many}; the same momentum reconstruction has been demonstrated at the single-atom level~\cite{brown2023time}. Separating internal states during expansion yields a spin-resolved momentum distribution~\cite{ValdesCuriel2021}.
The kinetic equation relates this distribution to the momentum-resolved source kernels \(\mathcal K_a^X(\vb r,\vb k)\).
The source \(\mathcal S_a^X\) appearing in Eq.~\eqref{eq:spinContinuity} is the unweighted momentum integral \(\mathcal S_a^X=\intk\mathcal K_a^X\).
The spin-current moment instead contains \(\intk v_i\mathcal K_a^X\), while higher moments and detector projections use the momentum factors appropriate to each observable.
For scalar input, \(\mathcal K_a^{\Lambda(1)}\) is the coherent first-gradient kernel, whereas \(\mathcal K_a^{\Gamma(2)}\) is the leading gradient-specific dissipative kernel of the local Markov model; they enter at first and second Moyal order, respectively.
Keeping the kernels explicit preserves their momentum harmonics before the detector weight and momentum integration are applied.
Momentum averaging is imposed only after this separation, when the selected weight determines which spin observable survives the clean-continuum parity constraints.
In the ballistic far-field limit, a protocol with a spin- and momentum-independent real-space selection function \(A(\vb r)\), applied before release without distorting the selected momentum distribution, measures
\begin{equation}
n_a^{\mathrm{TOF}}(\vb k;A)=\int d\vb r\,A(\vb r)f_a(\vb r,\vb k).
\label{eq:windowedTOF}
\end{equation}
Conventional whole-cloud imaging corresponds to \(A=1\). In the local-response limit with a spatially uniform scalar drive and a spatially uniform proportionality between the generated spin distribution and its source kernel, the curvature contribution factorizes into its displayed spatial and momentum derivatives. Two integrations by parts of its spatial factor yield
\begin{equation}
\int d\vb r\,A\,\partial_i\partial_j\Gamma_a
=\int d\vb r\,\Gamma_a\,\partial_i\partial_j A
+\text{boundary terms}.
\label{eq:spatialWindow}
\end{equation}
Thus, within this factorized idealization, a localized profile with vanishing boundary gradients cancels in a whole-cloud image. A nonuniform selector, spatial variation of the scalar drive, depletion, or effective relaxation, or boundary conversion can retain the local mixed-curvature signal and requires protocol-specific modeling. For a selector of width \(L_A\), local response and isolation of the curvature region require \(\ell_J\ll L_A\lesssim L_{\rm grad}\), where \(\ell_J\) is the current-relaxation length and \(L_{\rm grad}\) is the programmed-profile scale. Restoring \(\hbar\), finite initial extent and detector resolution yield momentum uncertainties \(mL_A/(\hbar t_{\rm TOF})\) and \(m\delta x/(\hbar t_{\rm TOF})\), respectively; both should be smaller than the drive-induced scale \(F_x\tau_{\rm tr}/\hbar\). The protocol requires these inequalities; numerical contrast remains specific to the chosen species and selection protocol.

\paragraph{Coherent-Gradient Spin Current---}
We first examine how the coherent self-energy gradient
\(\partial_j\Lambda_a\) generates a spin current.
This coherent channel is a direct first-gradient conversion result and the comparison baseline; the central result of this Letter is the dissipative-curvature source analyzed afterward.
Taking ${\rm Tr}\{\sigma_av_i\cdots\}$ of Eq.~\eqref{eq:qkt}, we obtain
\begin{align}
\partial_t j_{s,i}^{a}+\partial_j\Pi_{ij}^{a}
&\simeq
Q_i^{a,\Lambda(1)}
+Q_i^{a,\Gamma(1)}
+Q_i^{a,\Gamma(2)}
\nonumber\\
&{}
+2\epsilon_{abc}\Lambda_b j_{s,i}^{c}
-\frac{j_{s,i}^{a}}{\tau_J},
\label{eq:spinCurrentMoment}
\end{align}
where \(\Pi_{ij}^{a}=\intk v_iv_jf_a\) and \(Q_i^{a,X}=\intk v_i\mathcal K_a^X\).
For a locally isotropic reference gas, meaning that the dispersion and linearized collision kernel are invariant under momentum rotations at each \(\vb r\), the angular dependence of \(v_i\) lies in the \(\ell=1\) sector. Orthogonality of angular harmonics then restricts \(j_{s,i}^{a}=\intk v_if_a\) to the \(\ell=1\), velocity-odd component of \(f_a(\vb k)\).
Within a single-relaxation-time closure, we approximate the collision term of this harmonic as \((\partial_t f_a^{(\ell=1)})_{\rm coll}=-f_a^{(\ell=1)}/\tau_J\). Multiplication by \(v_i\) followed by momentum integration yields the last term in Eq.~\eqref{eq:spinCurrentMoment}; \(\tau_J\) therefore relaxes the current-carrying momentum anisotropy rather than a uniform spin density.
The programmed profile breaks real-space isotropy and makes the driven distribution anisotropic; this remains compatible with the local momentum-space closure when the profile varies on length scales much longer than the local mean free path characterized by the current-relaxation time \(\tau_J\). If the local dispersion or collision kernel is anisotropic, angular harmonics generally mix; a quantitative treatment replaces the scalar \(\tau_J\) by a position-dependent tensor or a multimode collision operator.
For the coherent scalar-to-spin term,
\begin{align}
Q_i^{a,\Lambda(1)}
&=
(\partial_j\Lambda_a)\intk v_i\partial_{k_j}f_0=-\chi_{ij}^{(v)}\partial_j\Lambda_a.
\label{eq:QLambda}
\end{align}
Here \(\chi_{ij}^{(v)}=\intk f_0\,\partial_{k_j}v_i\).
For a locally isotropic gas with a parabolic dispersion, \(\chi_{ij}^{(v)}=(n/m)\delta_{ij}\).
The coherent gradient therefore acts as a spin-dependent force that injects the velocity-odd component of \(f_a(\vb k)\) entering \(j_{s,i}^a\), even when spin relaxation is absent.

\paragraph{Coherent-Gradient Spin Chirality---}
The term \(Q_i^{a,\Lambda(1)}\) in Eq.~\eqref{eq:QLambda} has the structure of a spin-dependent force that injects a velocity-odd spin distribution.
For a collinear texture \(\vb\Lambda(z)=\Lambda(z)\hat{\vb n}\) whose spin direction \(\hat{\vb n}\) is independent of \(z\), the induced spin current is parallel to \(\hat{\vb n}\); hence the local coherent torque \(2\epsilon_{abc}\Lambda_b j_{s,i}^{c}\) in Eq.~\eqref{eq:spinCurrentMoment} vanishes.
By contrast, for a general noncollinear texture, the torque acts on the gradient-injected current and generates a chirality contribution.
Solving Eq.~\eqref{eq:spinCurrentMoment} iteratively in the open, slowly varying limit yields
\begin{align}
j_{s,i}^{a}
&\simeq
-\tau_J\chi_{ij}^{(v)}\partial_j\Lambda_a
-2\tau_J^2\chi_{ij}^{(v)}
\epsilon_{abc}\Lambda_b\partial_j\Lambda_c,
\label{eq:lambdaChirality}
\end{align}
which holds for a general texture; in the collinear limit only the first term survives, whereas the noncollinear chirality enters through the second term.
The second term is a coherent-gradient chirality current generated by the spin-space chirality \((\vb\Lambda\times\partial_j\vb\Lambda)_a\), with the structure of a Leggett-Rice contribution~\cite{PhysRevLett.20.586}.
Whereas the Leggett-Rice effect~\cite{Trotzky2015} arises from interaction-induced spin precession, here the same chirality is produced by a programmed static texture \(\vb\Lambda(\vb r)\), making it a controllable self-energy-gradient channel rather than an interaction effect.
It is one of the direct spin-conversion channels produced by programmable self-energy gradients.

\paragraph{Dissipative-Curvature Spin Response---}
We now consider scalar-to-spin conversion mediated by the dissipative texture \(\vb\Gamma(\vb r)\).
As the first step, a spin-independent force \(F_x\) along \(x\) prepares a nonequilibrium scalar distribution
\begin{align}
\delta f_0^{(F_x)}(\vb k)
&=
F_x\tau_{\mathrm{tr}}v_x
\left(-\partial_{\varepsilon}f_{\rm eq}\right),
\label{eq:deltaFx}
\end{align}
where \(\tau_{\rm tr}\) is the scalar transport time
and atoms accumulate at \(v_x>0\).
For an unpolarized reference environment, \(\Delta f_0^\Sigma=\delta f_0^{(F_x)}\).
For the force-driven scalar input, \(\Delta f^\Sigma=\delta f_0^{(F_x)}\one\), and hence \(\Delta f_c^\Sigma=0\).
Equation~\eqref{eq:kernelGamma} therefore implies \(\mathcal K_a^{\Gamma(1)}=(1/2)\epsilon_{abc}(\partial_i\Gamma_b)\partial_{k_i}\Delta f_c^\Sigma=0\) at each momentum. In matrix form, the corresponding first-gradient term is a spin-space commutator with the scalar matrix \(\delta f_0^{(F_x)}\one\); it vanishes because \(\one\) commutes with every spin matrix.
This pointwise vanishing follows from spin algebra rather than momentum integration. Within the local Markov model, the leading gradient-specific dissipative scalar-to-spin source is therefore the second-order curvature term, whose relevant components are
\begin{align}
\mathcal K_{a,zz}^{\Gamma(2)}(\vb r,\vb k)
&=
\frac{1}{8}
\left(\partial_z^2\Gamma_a\right)
\partial_{k_z}^2\delta f_0^{(F_x)},
\label{eq:Sdir}\\
\mathcal K_{a,xz}^{\Gamma(2)}(\vb r,\vb k)
&=
\frac{1}{4}
\left(\partial_x\partial_z\Gamma_a\right)
\partial_{k_x}\partial_{k_z}\delta f_0^{(F_x)} .
\label{eq:SdirX}
\end{align}
The factor \(1/4\) in Eq.~\eqref{eq:SdirX} in the mixed-curvature term follows because the symmetric tensor sum in Eq.~\eqref{eq:kernelGammaSecond} contains both the \(xz\) and \(zx\) components, each with coefficient \(1/8\).
While Eq.~\eqref{eq:Sdir} is \(k_x\)-odd and \(k_z\)-even for the force-driven distribution in Eq.~\eqref{eq:deltaFx},
the mixed-curvature source in Eq.~\eqref{eq:SdirX} is \(k_x\)-even and \(k_z\)-odd.
These two curvature components therefore produce complementary local spin textures: the \(zz\) curvature produces a longitudinally driven spin texture, whereas the \(xz\) curvature carries the mixed longitudinal-transverse parity needed for transverse readout.
For a specified \(\vb r\), these textures remain finite before momentum averaging.  A spin-resolved time-of-flight protocol that selects the curvature region can test their reversal under \(F_x\) or the programmed curvature and their disappearance for a uniform profile; whole-cloud imaging alone need not retain the mixed-derivative signal.
Whether the same source survives in an integrated spin current or spin-density moment is decided by the detector weight, through the selection rule given in the End Matter.
The spin-density moment
\begin{equation}
D_z^a
=
\frac{1}{N}\int d\vb r\,z\,s_a(\vb r)
\label{eq:readout}
\end{equation}
is one such finite-geometry observable.
$D_z^a$ quantifies the real-space spin separation illustrated in Fig.~\ref{fig:concept}, and its value follows from the momentum-space texture only once propagation, geometry, boundary conditions, and the detector protocol are specified, as detailed in the End Matter.
In a cold-atom cloud, the spin current is the transport quantity produced by the kinetic equation, whereas \(D_z^a\) and a spatially selected spin-resolved time-of-flight texture are experimental readouts with distinct propagation and detector requirements.
The static finite-geometry weight need not transform under drive reversal. Writing \(\mathcal G_a=\partial_i\partial_j\Gamma_a\) for the programmed curvature component, define the drive-antisymmetrized dipole
\begin{equation}
D_{z,{\rm odd}}^a(F_x,\mathcal G_a)
=\frac{D_z^a(F_x,\mathcal G_a)-D_z^a(-F_x,\mathcal G_a)}{2}.
\label{eq:oddDipole}
\end{equation}
This combination removes \(F_x\)-even backgrounds, including the equilibrium coherent-gradient response. In the absence of other \(F_x\)-odd sources, its dissipative-curvature component obeys
\begin{align}
D_{z,{\rm odd}}^a(F_x,\mathcal G_a)
&=
-D_{z,{\rm odd}}^a(-F_x,\mathcal G_a),
\nonumber\\
D_{z,{\rm odd}}^a(F_x,\mathcal G_a)
&=
-D_{z,{\rm odd}}^a(F_x,-\mathcal G_a),
\nonumber\\
D_{z,{\rm odd}}^a(F_x,0)
&=
0 .
\label{eq:tests}
\end{align}
The last two relations require reversal or removal of the complete programmed curvature profile and the absence of curvature-even, \(F_x\)-odd channels. The difference between opposite drive directions therefore isolates the dissipative-curvature contribution before the curvature-reversal test is applied.

\paragraph{Conclusion---}
We have shown that coherent self-energy gradients and dissipative self-energy curvature generate distinct spin-source structures in a two-component ultracold gas.
Starting from the quantum kinetic equation, we have obtained a spin-continuity equation in which coherent first gradients and dissipative curvature terms enter as spin
source structures.
The coherent gradient acts as a spin-dependent force and produces a Leggett-Rice-type spin-current chirality for noncollinear textures.
The dissipative curvature converts a spin-independent drive into local spin textures that can be accessed by source-region-selected spin-resolved time-of-flight imaging, real-space apertures, boundary conversion, or spin-density moments.
In the End Matter we show the connection between this source/readout separation and gradient-material spin conversion, where direct and inverse measurements project different source amplitudes, propagation lengths, and boundary weights in finite films~\cite{Okano2019,Horaguchi2025}.
A complete analog of finite-device gradient materials requires an optical-lattice dispersion or another nonlinear momentum weight; the present continuum gas establishes the local curvature kernel and a spatially selected time-of-flight protocol for testing its sign reversals.
Thus real-space self-energy profiles provide a controllable spin source for programmable gradient quantum matter, with the spatially windowed, parity-locked momentum texture as its primary falsifiable signature.

\begin{acknowledgments}
M.M. thanks Y. Nozaki for valuable discussions on gradient-material spintronics, which helped motivate the present cold-atom analog perspective.
This work was supported by the National Natural Science Foundation of China (NSFC) under Grant No. 12374126, by the Priority Program of Chinese Academy of Sciences under Grant No. XDB28000000, and by Japan Society for the Promotion of Science (JSPS) Grants-in-Aid for Scientific Research (KAKENHI) Grant Nos. JP22K13981, JP23H01839, JP23K22429, JP24H00322, JP24K06951, JP25K17351, and JP26K07063.
Y.S. is supported by the RIKEN TRIP initiative (RIKEN Quantum).
\end{acknowledgments}

\bibliography{ref}

@article{Lin2011Science,
  author = {Lin, Y.-J. and Jimenez-Garcia, K. and Spielman, I. B.},
  title = {Spin-orbit-coupled Bose-Einstein condensates},
  journal = {Nature},
  volume = {471},
  pages = {83--86},
  year = {2011},
  doi = {10.1038/nature09887}
}

@article{Zhai2015Review,
  author = {Zhai, Hui},
  title = {Degenerate quantum gases with spin-orbit coupling: a review},
  journal = {Reports on Progress in Physics},
  volume = {78},
  number = {2},
  pages = {026001},
  year = {2015},
  doi = {10.1088/0034-4885/78/2/026001}
}

@article{Trotzky2015,
  author = {Trotzky, S. and Beattie, S. and Luciuk, C. and Smale, S. and Bardon, A. B. and Enss, T. and Taylor, E. and Zhang, S. and Thywissen, J. H.},
  title = {Observation of the Leggett-Rice effect in a unitary Fermi gas},
  journal = {Physical Review Letters},
  volume = {114},
  number = {1},
  pages = {015301},
  year = {2015},
  doi = {10.1103/PhysRevLett.114.015301}
}

@article{PhysRevLett.20.586,
  title = {Spin Echoes in Liquid ${\mathrm{He}}^{3}$ and Mixtures: A Predicted New Effect},
  author = {Leggett, A. J. and Rice, M. J.},
  journal = {Phys. Rev. Lett.},
  volume = {20},
  issue = {12},
  pages = {586--589},
  numpages = {0},
  year = {1968},
  month = {Mar},
  publisher = {American Physical Society},
  doi = {10.1103/PhysRevLett.20.586},
  url = {https://link.aps.org/doi/10.1103/PhysRevLett.20.586}
}

@article{Royse2026,
  author = {Royse, Camen A. and Thomas, J. E.},
  title = {Chirality-Induced Spin Currents in a Fermi Gas},
  journal = {Physical Review Letters},
  volume = {136},
  number = {12},
  pages = {123401},
  year = {2026},
  doi = {10.1103/73th-7hwt}
}

@article{Corman2019PRA,
  author = {Corman, L. and Fabritius, P. and Haeusler, S. and Mohan, J. and Dogra, L. H. and Husmann, D. and Lebrat, M. and Esslinger, T.},
  title = {Quantized conductance through a dissipative atomic point contact},
  journal = {Physical Review A},
  volume = {100},
  number = {5},
  pages = {053605},
  year = {2019},
  doi = {10.1103/PhysRevA.100.053605}
}

@article{Sinova2015,
  author = {Sinova, Jairo and Valenzuela, Sergio O. and Wunderlich, J. and Back, C. H. and Jungwirth, T.},
  title = {Spin Hall effects},
  journal = {Reviews of Modern Physics},
  volume = {87},
  number = {4},
  pages = {1213--1260},
  year = {2015},
  doi = {10.1103/RevModPhys.87.1213}
}

@article{Manchon2019,
  author = {Manchon, Aur{\'e}lien and {\v Z}elezn{\'y}, Jaroslav and Miron, Ioan M. and Jungwirth, Tom{\'a}{\v s} and Sinova, Jairo and Thiaville, Andr{\'e} and Garello, Kevin and Gambardella, Pietro},
  title = {Current-induced spin-orbit torques in ferromagnetic and antiferromagnetic systems},
  journal = {Reviews of Modern Physics},
  volume = {91},
  number = {3},
  pages = {035004},
  year = {2019},
  doi = {10.1103/RevModPhys.91.035004}
}

@article{An2016NatCommun,
  author = {An, Hongyu and Kageyama, Yuito and Kanno, Yusuke and Enishi, Nagisa and Ando, Kazuya},
  title = {Spin-torque generator engineered by natural oxidation of Cu},
  journal = {Nature Communications},
  volume = {7},
  pages = {13069},
  year = {2016},
  doi = {10.1038/ncomms13069}
}

@article{Nakayama2023,
  author = {Nakayama, Hayato and Horaguchi, Taisuke and He, Cong and Sukegawa, Hiroaki and Ohkubo, Tadakatsu and Mitani, Seiji and Yamanoi, Kazuto and Nozaki, Yukio},
  title = {Spin-torque generation using a compositional gradient at the interface between titanium and tungsten thin films},
  journal = {Physical Review B},
  volume = {107},
  number = {17},
  pages = {174416},
  year = {2023},
  doi = {10.1103/PhysRevB.107.174416}
}

@article{Okano2019,
  author = {Okano, Genki and Matsuo, Mamoru and Ohnuma, Yuichi and Maekawa, Sadamichi and Nozaki, Yukio},
  title = {Nonreciprocal spin current generation in surface-oxidized copper films},
  journal = {Physical Review Letters},
  volume = {122},
  number = {21},
  pages = {217701},
  year = {2019},
  doi = {10.1103/PhysRevLett.122.217701}
}

@article{Horaguchi2025,
  author = {Horaguchi, Taisuke and He, Cong and Wen, Zhenchao and Nakayama, Hayato and Ohkubo, Tadakatsu and Mitani, Seiji and Sukegawa, Hiroaki and Fujimoto, Junji and Yamanoi, Kazuto and Matsuo, Mamoru and Nozaki, Yukio},
  title = {Nanometer-thick Si/Al gradient materials for spin torque generation},
  journal = {Science Advances},
  volume = {11},
  number = {19},
  pages = {eadr9481},
  year = {2025},
  doi = {10.1126/sciadv.adr9481}
}

@article{Shi2026AFM,
  author = {Shi, Yanan and Huang, Qikun and Yi, Longwen and Han, Zhen and Tan, Cheng and Han, Xiang and Cai, Li and Cao, Qiang and Tian, Yufeng and Yan, Shishen},
  title = {Efficient Spin--Orbit Torque Switching via Spin--Vorticity Coupling Effect in Light-Metal/Ferromagnetic-Metal Bilayers},
  journal = {Advanced Functional Materials},
  pages = {e76194},
  year = {2026},
  doi = {10.1002/adfm.76194}
}

@article{Yi2025PRL,
  author = {Yi, Longwen and Yang, Tianxiang and Tan, Cheng and Xie, Ronghuan and Liu, Senmiao and Cai, Li and Cao, Qiang and Wang, Yan and L{\"u}, Weiming and Tian, Yufeng and Huang, QiKun and Yan, Shishen},
  title = {Large Orbital Torque from Interfacial Spin-Vorticity Coupling in PtCo/Cu Heterostructures},
  journal = {Physical Review Letters},
  volume = {135},
  number = {15},
  pages = {156702},
  year = {2025},
  doi = {10.1103/qgdy-k39l}
}

@article{Kontani2009PRL,
  author = {Kontani, H. and Tanaka, T. and Hirashima, D. S. and Yamada, K. and Inoue, J.},
  title = {Giant Orbital Hall Effect in Transition Metals: Origin of Large Spin and Anomalous Hall Effects},
  journal = {Physical Review Letters},
  volume = {102},
  number = {1},
  pages = {016601},
  year = {2009},
  doi = {10.1103/PhysRevLett.102.016601}
}

@article{GoJoKimLee2018PRL,
  author = {Go, Dongwook and Jo, Daegeun and Kim, Changyoung and Lee, Hyun-Woo},
  title = {Intrinsic Spin and Orbital Hall Effects from Orbital Texture},
  journal = {Physical Review Letters},
  volume = {121},
  number = {8},
  pages = {086602},
  year = {2018},
  doi = {10.1103/PhysRevLett.121.086602}
}

@article{JoGoLee2018PRB,
  author = {Jo, Daegeun and Go, Dongwook and Lee, Hyun-Woo},
  title = {Gigantic intrinsic orbital Hall effects in weakly spin-orbit coupled metals},
  journal = {Physical Review B},
  volume = {98},
  number = {21},
  pages = {214405},
  year = {2018},
  doi = {10.1103/PhysRevB.98.214405}
}

@article{GoLee2020PRR,
  author = {Go, Dongwook and Lee, Hyun-Woo},
  title = {Orbital torque: Torque generation by orbital current injection},
  journal = {Physical Review Research},
  volume = {2},
  number = {1},
  pages = {013177},
  year = {2020},
  doi = {10.1103/PhysRevResearch.2.013177}
}

@article{Go2020PRR,
  author = {Go, Dongwook and Freimuth, Frank and Hanke, Jan-Philipp and Xue, Fei and Gomonay, Olena and Lee, Kyung-Jin and Bl{\"u}gel, Stefan and Haney, Paul M. and Lee, Hyun-Woo and Mokrousov, Yuriy},
  title = {Theory of current-induced angular momentum transfer dynamics in spin-orbit coupled systems},
  journal = {Physical Review Research},
  volume = {2},
  number = {3},
  pages = {033401},
  year = {2020},
  doi = {10.1103/PhysRevResearch.2.033401}
}

@article{ValetRaimondi2025PRB,
  author = {Valet, T. and Raimondi, R.},
  title = {Quantum kinetic theory of the linear response for weakly disordered multiband systems},
  journal = {Physical Review B},
  volume = {111},
  number = {4},
  pages = {L041118},
  year = {2025},
  doi = {10.1103/PhysRevB.111.L041118}
}

@article{Valet2025PRL,
  author = {Valet, Thierry and Jaffr{\`e}s, Henri and Cros, Vincent and Raimondi, Roberto},
  title = {Quantum Kinetic Anatomy of Electron Angular Momenta Edge Accumulation},
  journal = {Physical Review Letters},
  volume = {135},
  number = {25},
  pages = {256301},
  year = {2025},
  doi = {10.1103/86cm-yn9z}
}

@book{KadanoffBaym1962,
  author = {Kadanoff, L. P. and Baym, G.},
  title = {Quantum Statistical Mechanics},
  publisher = {W. A. Benjamin},
  address = {New York},
  year = {1962}
}

@article{Keldysh1965,
  author = {Keldysh, L. V.},
  title = {Diagram technique for nonequilibrium processes},
  journal = {Soviet Physics JETP},
  volume = {20},
  pages = {1018--1026},
  year = {1965}
}

@article{RammerSmith1986,
  author = {Rammer, J. and Smith, H.},
  title = {Quantum field-theoretical methods in transport theory of metals},
  journal = {Reviews of Modern Physics},
  volume = {58},
  number = {2},
  pages = {323--359},
  year = {1986},
  doi = {10.1103/RevModPhys.58.323}
}

@book{HaugJauho2008,
  author = {Haug, Hartmut and Jauho, Antti-Pekka},
  title = {Quantum Kinetics in Transport and Optics of Semiconductors},
  edition = {2},
  series = {Springer Series in Solid-State Sciences},
  volume = {123},
  publisher = {Springer},
  address = {Berlin},
  year = {2008},
  doi = {10.1007/978-3-540-73564-9}
}

@article{PhysRevLett.122.040405,
  title = {Designer Spatial Control of Interactions in Ultracold Gases},
  author = {Arunkumar, N. and Jagannathan, A. and Thomas, J. E.},
  journal = {Phys. Rev. Lett.},
  volume = {122},
  issue = {4},
  pages = {040405},
  numpages = {6},
  year = {2019},
  month = {Feb},
  publisher = {American Physical Society},
  doi = {10.1103/PhysRevLett.122.040405},
  url = {https://link.aps.org/doi/10.1103/PhysRevLett.122.040405}
}

@article{PhysRevA.106.023322,
  title = {Realization of space-dependent interactions by an optically controlled magnetic $p$-wave Feshbach resonance in degenerate Fermi gases},
  author = {Bian, Guoqi and Huang, Lianghui and Li, Donghao and Meng, Zengming and Chen, Liangchao and Wang, Pengjun and Zhang, Jing},
  journal = {Phys. Rev. A},
  volume = {106},
  issue = {2},
  pages = {023322},
  numpages = {6},
  year = {2022},
  month = {Aug},
  publisher = {American Physical Society},
  doi = {10.1103/PhysRevA.106.023322},
  url = {https://link.aps.org/doi/10.1103/PhysRevA.106.023322}
}

@article{PhysRevLett.101.150401,
  title = {Observation of Anomalous Spin Segregation in a Trapped Fermi Gas},
  author = {Du, X. and Luo, L. and Clancy, B. and Thomas, J. E.},
  journal = {Phys. Rev. Lett.},
  volume = {101},
  issue = {15},
  pages = {150401},
  numpages = {4},
  year = {2008},
  month = {Oct},
  publisher = {American Physical Society},
  doi = {10.1103/PhysRevLett.101.150401},
  url = {https://link.aps.org/doi/10.1103/PhysRevLett.101.150401}
}

@article{PhysRevLett.103.010401,
  title = {Controlling Spin Current in a Trapped Fermi Gas},
  author = {Du, X. and Zhang, Y. and Petricka, J. and Thomas, J. E.},
  journal = {Phys. Rev. Lett.},
  volume = {103},
  issue = {1},
  pages = {010401},
  numpages = {4},
  year = {2009},
  month = {Jul},
  publisher = {American Physical Society},
  doi = {10.1103/PhysRevLett.103.010401},
  url = {https://link.aps.org/doi/10.1103/PhysRevLett.103.010401}
}

@article{sommer2011universal,
  title={Universal spin transport in a strongly interacting Fermi gas},
  author={Sommer, Ariel and Ku, Mark and Roati, Giacomo and Zwierlein, Martin W},
  journal={Nature},
  volume={472},
  number={7342},
  pages={201--204},
  year={2011},
  publisher={Nature Publishing Group UK London}
}

@article{koschorreck2013universal,
  title={Universal spin dynamics in two-dimensional Fermi gases},
  author={Koschorreck, Marco and Pertot, Daniel and Vogt, Enrico and K{\"o}hl, Michael},
  journal={Nature Physics},
  volume={9},
  number={7},
  pages={405--409},
  year={2013},
  publisher={Nature Publishing Group UK London}
}

@article{sommer2011spin,
  title={Spin transport in polaronic and superfluid Fermi gases},
  author={Sommer, Ariel and Ku, Mark and Zwierlein, Martin W},
  journal={New Journal of Physics},
  volume={13},
  number={5},
  pages={055009},
  year={2011}
}

@article{valtolina2017exploring,
  title={Exploring the ferromagnetic behaviour of a repulsive Fermi gas through spin dynamics},
  author={Valtolina, G and Scazza, F and Amico, A and Burchianti, A and Recati, A and Enss, T and Inguscio, M and Zaccanti, M and Roati, G},
  journal={Nature Physics},
  volume={13},
  number={7},
  pages={704--709},
  year={2017},
  publisher={Nature Publishing Group UK London}
}

@article{bardon2014transverse,
  title={Transverse demagnetization dynamics of a unitary Fermi gas},
  author={Bardon, AB and Beattie, S and Luciuk, C and Cairncross, W and Fine, D and Cheng, NS and Edge, GJA and Taylor, E and Zhang, Shizhong and Trotzky, S and others},
  journal={Science},
  volume={344},
  number={6185},
  pages={722--724},
  year={2014},
  publisher={American Association for the Advancement of Science},
  doi={10.1126/science.1247425}
}

@article{krinner2016mapping,
  title={Mapping out spin and particle conductances in a quantum point contact},
  author={Krinner, Sebastian and Lebrat, Martin and Husmann, Dominik and Grenier, Charles and Brantut, Jean-Philippe and Esslinger, Tilman},
  journal={Proceedings of the National Academy of Sciences},
  volume={113},
  number={29},
  pages={8144--8149},
  year={2016},
  publisher={National Academy of Sciences}
}

@article{9ks8-zv9b,
  title = {Saturation of Thermal and Spin Conductances in a Dissipative Superfluid Junction},
  author = {Huang, Meng-Zi and Fabritius, Philipp and Mohan, Jeffrey and Talebi, Mohsen and Wili, Simon and Esslinger, Tilman},
  journal = {Phys. Rev. Lett.},
  volume = {134},
  issue = {25},
  pages = {253403},
  numpages = {7},
  year = {2025},
  month = {Jun},
  publisher = {American Physical Society},
  doi = {10.1103/9ks8-zv9b},
  url = {https://link.aps.org/doi/10.1103/9ks8-zv9b}
}

@article{PhysRevLett.118.130405,
  title = {Observation of Quantum-Limited Spin Transport in Strongly Interacting Two-Dimensional Fermi Gases},
  author = {Luciuk, C. and Smale, S. and B\"ottcher, F. and Sharum, H. and Olsen, B. A. and Trotzky, S. and Enss, T. and Thywissen, J. H.},
  journal = {Phys. Rev. Lett.},
  volume = {118},
  issue = {13},
  pages = {130405},
  numpages = {6},
  year = {2017},
  month = {Mar},
  publisher = {American Physical Society},
  doi = {10.1103/PhysRevLett.118.130405},
  url = {https://link.aps.org/doi/10.1103/PhysRevLett.118.130405}
}

@article{PhysRevLett.109.050403,
  title = {Dynamic Spin Response of a Strongly Interacting Fermi Gas},
  author = {Hoinka, S. and Lingham, M. and Delehaye, M. and Vale, C. J.},
  journal = {Phys. Rev. Lett.},
  volume = {109},
  issue = {5},
  pages = {050403},
  numpages = {5},
  year = {2012},
  month = {Aug},
  publisher = {American Physical Society},
  doi = {10.1103/PhysRevLett.109.050403},
  url = {https://link.aps.org/doi/10.1103/PhysRevLett.109.050403}
}

@article{krauser2014giant,
  title={Giant spin oscillations in an ultracold Fermi sea},
  author={Krauser, Jasper Simon and Ebling, Ulrich and Fl{\"a}schner, Nick and Heinze, Jannes and Sengstock, Klaus and Lewenstein, Maciej and Eckardt, Andr{\'e} and Becker, Christoph},
  journal={Science},
  volume={343},
  number={6167},
  pages={157--160},
  year={2014},
  publisher={American Association for the Advancement of Science}
}

@article{PhysRevA.108.L041304,
  title = {Verifying a quasiclassical spin model of perturbed quantum rewinding in a Fermi gas},
  author = {Huang, J. and Royse, Camen A. and Arakelyan, I. and Thomas, J. E.},
  journal = {Phys. Rev. A},
  volume = {108},
  issue = {4},
  pages = {L041304},
  numpages = {5},
  year = {2023},
  month = {Oct},
  publisher = {American Physical Society},
  doi = {10.1103/PhysRevA.108.L041304},
  url = {https://link.aps.org/doi/10.1103/PhysRevA.108.L041304}
}

@article{PhysRevA.99.063620,
  title = {Spin-energy correlation in degenerate weakly interacting Fermi gases},
  author = {Pegahan, S. and Kangara, J. and Arakelyan, I. and Thomas, J. E.},
  journal = {Phys. Rev. A},
  volume = {99},
  issue = {6},
  pages = {063620},
  numpages = {14},
  year = {2019},
  month = {Jun},
  publisher = {American Physical Society},
  doi = {10.1103/PhysRevA.99.063620},
  url = {https://link.aps.org/doi/10.1103/PhysRevA.99.063620}
}

@incollection{grimm2000optical,
  title={Optical dipole traps for neutral atoms},
  author={Grimm, Rudolf and Weidem{\"u}ller, Matthias and Ovchinnikov, Yurii B},
  booktitle={Advances in atomic, molecular, and optical physics},
  volume={42},
  pages={95--170},
  year={2000},
  publisher={Elsevier}
}

@article{PhysRevLett.115.073002,
  title = {Creating State-Dependent Lattices for Ultracold Fermions by Magnetic Gradient Modulation},
  author = {Jotzu, Gregor and Messer, Michael and G\"org, Frederik and Greif, Daniel and Desbuquois, R\'emi and Esslinger, Tilman},
  journal = {Phys. Rev. Lett.},
  volume = {115},
  issue = {7},
  pages = {073002},
  numpages = {5},
  year = {2015},
  month = {Aug},
  publisher = {American Physical Society},
  doi = {10.1103/PhysRevLett.115.073002},
  url = {https://link.aps.org/doi/10.1103/PhysRevLett.115.073002}
}

@article{nichols2019spin,
  title={Spin transport in a Mott insulator of ultracold fermions},
  author={Nichols, Matthew A and Cheuk, Lawrence W and Okan, Melih and Hartke, Thomas R and Mendez, Enrique and Senthil, T and Khatami, Ehsan and Zhang, Hao and Zwierlein, Martin W},
  journal={Science},
  volume={363},
  number={6425},
  pages={383--387},
  year={2019},
  publisher={American Association for the Advancement of Science}
}

@article{gauthier2016direct,
  title={Direct imaging of a digital-micromirror device for configurable microscopic optical potentials},
  author={Gauthier, G and Lenton, I and McKay Parry, N and Baker, M and Davis, MJ and Rubinsztein-Dunlop, H and Neely, TW},
  journal={Optica},
  volume={3},
  number={10},
  pages={1136--1143},
  year={2016},
  publisher={Optical Society of America}
}

@article{ren2022chiral,
  title={Chiral control of quantum states in non-Hermitian spin--orbit-coupled fermions},
  author={Ren, Zejian and Liu, Dong and Zhao, Entong and He, Chengdong and Pak, Ka Kwan and Li, Jensen and Jo, Gyu-Boong},
  journal={Nature Physics},
  volume={18},
  number={4},
  pages={385--389},
  year={2022},
  publisher={Nature Publishing Group UK London},
  doi={10.1038/s41567-021-01491-x}
}

@article{zhao2025two,
  title={Two-dimensional non-Hermitian skin effect in an ultracold Fermi gas},
  author={Zhao, Entong and Wang, Zhiyuan and He, Chengdong and Poon, Ting Fung Jeffrey and Pak, Ka Kwan and Liu, Yu-Jun and Ren, Peng and Liu, Xiong-Jun and Jo, Gyu-Boong},
  journal={Nature},
  volume={637},
  number={8046},
  pages={565--573},
  year={2025},
  publisher={Nature Publishing Group UK London}
}

@article{brown2023time,
  title={Time-of-flight quantum tomography of an atom in an optical tweezer},
  author={Brown, M. O. and Muleady, S. R. and Dworschack, W. J. and Lewis-Swan, R. J. and Rey, A. M. and Romero-Isart, O. and Regal, C. A.},
  journal={Nature Physics},
  volume={19},
  number={4},
  pages={569--573},
  year={2023},
  doi={10.1038/s41567-022-01890-8}
}

@article{Bloch2008many,
  title={Many-body physics with ultracold gases},
  author={Bloch, Immanuel and Dalibard, Jean and Zwerger, Wilhelm},
  journal={Reviews of Modern Physics},
  volume={80},
  pages={885--964},
  year={2008},
  doi={10.1103/RevModPhys.80.885}
}

@article{ValdesCuriel2021,
  title={Topological features without a lattice in {Rashba} spin-orbit coupled atoms},
  author={Vald{\'e}s-Curiel, A. and Trypogeorgos, D. and Liang, Q.-Y. and Anderson, R. P. and Spielman, I. B.},
  journal={Nature Communications},
  volume={12},
  pages={593},
  year={2021},
  doi={10.1038/s41467-020-20762-4}
}

\section*{End Matter}

\paragraph{Second-Order Source Derivation---}
Equation~\eqref{eq:spinProjected} follows from the Moyal expansion in Eq.~\eqref{eq:expandedCollision}.
For scalar input, \(\Delta f^\Sigma=\Delta f_0^\Sigma\one\), the dissipative first-gradient term is
\begin{align}
-\frac{i}{4}
\left[
\partial_i\Gamma_a\sigma_a,
\partial_{k_i}\Delta f_0^\Sigma\one
\right]
&=
0 .
\label{eq:end_gamma_first_zero}
\end{align}
The leading dissipative scalar-to-spin term is therefore the second-gradient anticommutator,
\begin{align}
\frac{1}{16}
\left\{
\partial_i\partial_j\Gamma_a\sigma_a,
\partial_{k_i}\partial_{k_j}\Delta f_0^\Sigma\one
\right\}
&=
\frac{1}{8}
\left(\partial_i\partial_j\Gamma_a\right)
\nonumber\\
&\times
\partial_{k_i}\partial_{k_j}\Delta f_0^\Sigma
\sigma_a .
\label{eq:end_gamma_second_source}
\end{align}
Projection with \(\Tr(\sigma_b\sigma_a)/2=\delta_{ab}\) gives the first term of Eq.~\eqref{eq:kernelGammaSecond}.
By contrast, the coherent first-gradient term acting on a scalar distribution gives
\begin{align}
\frac{1}{2}
\left\{
\partial_i\Lambda_a\sigma_a,
\partial_{k_i}f_0\one
\right\}
&=
\left(\partial_i\Lambda_a\right)
\partial_{k_i}f_0\,\sigma_a ,
\label{eq:end_lambda_first_source}
\end{align}
which is Eq.~\eqref{eq:kernelLambda}.
This is the order distinction used in the main text: coherent scalar-to-spin conversion appears at first spatial-gradient order, while gradient-specific local Markov dissipative scalar-to-spin conversion appears at curvature order.
Uniform spin-selective loss belongs to the local term in Eq.~\eqref{eq:end_local_source} and should be distinguished from this spatial-conversion mechanism.

\paragraph{Momentum-Weight Selection---}
To compare the cold-atom source with gradient-material measurements, the relevant observables are momentum-weighted quantities fixed by the detector weight applied to the distribution.
We examine here how the gradient source projects onto the spin density, spin current, and higher moments.
A measurement with momentum weight \(W(\vb k)\) reads the spin distribution as
\begin{align}
O_W^a(\vb r)
&=
\intk W(\vb k) f_a(\vb r,\vb k),
\label{eq:weightedObservable}
\end{align}
with the spin density and spin current given by \(W=1\) and \(W=v_i\).
For a scalar nonequilibrium distribution, integration by parts gives the leading coherent and dissipative contributions
\begin{align}
\intk W\mathcal K_a^{\Lambda(1)}
&=
-\left(\partial_i\Lambda_a\right)
\intk f_0\,\partial_{k_i}W,
\label{eq:weightedLambda}\\
\intk W\mathcal K_a^{\Gamma(2)}
&=
\frac{1}{8}
\left(\partial_i\partial_j\Gamma_a\right)
\intk \Delta f_0^\Sigma\,
\partial_{k_i}\partial_{k_j}W ,
\label{eq:weightedGamma}
\end{align}
hence coherent gradients are selected by the first momentum derivative of \(W\) and the dissipative curvature by its second.
For the force-driven curvature source, the general detector-weight response is
\begin{align}
O_{W,\Gamma}^{a(2)}(\vb r)
&=
\intk W(\vb k)\mathcal K_a^{\Gamma(2)}
\nonumber\\
&=
\frac{1}{8}
\left(\partial_i\partial_j\Gamma_a\right)
\intk \delta f_0^{(F_x)}
\partial_{k_i}\partial_{k_j}W .
\label{eq:finiteDirect}
\end{align}
This gives three experimentally distinct projections:
(i) a spin density (\(W=1\)) removes pure momentum derivatives in the clean continuum;
(ii) a spin current (\(W=v_i\)) with a parabolic dispersion removes the local dissipative curvature because \(\partial_{k_j}\partial_{k_l}v_i=0\); and
(iii) finite apertures, nonparabolic optical-lattice dispersion, boundary conversion, or direct momentum-resolved imaging supply nonlinear weights that keep the curvature-generated texture.
The mixed-curvature kernel of Eq.~\eqref{eq:SdirX} therefore remains finite at each \((\vb r,\vb k)\).  Its momentum-integrated spin current requires such a nonlinear momentum weight, while its time-of-flight detection additionally requires the real-space selection in Eq.~\eqref{eq:windowedTOF} or a boundary contribution.

\paragraph{Local Spin Terms and Propagation---}
For the decomposition in Eq.~\eqref{eq:fdecomp}, the local part of the spin source entering Eq.~\eqref{eq:spinContinuity} may be written schematically as
\begin{align}
\mathcal S_a^{\mathrm{loc}}
&=
\intk
\left[
2\epsilon_{abc}\Lambda_b f_c
-\Gamma_0\Delta f_a^\Sigma
-\Gamma_a\Delta f_0^\Sigma
\right]
-\frac{s_a}{\tau_s^a}.
\label{eq:end_local_source}
\end{align}
The last term is a phenomenological spin-relaxation channel that can be omitted or replaced by a microscopic relaxation kernel in a more detailed model.
If finite-time propagation or boundary conversion is needed, the curvature-generated spin distribution can be represented as
\begin{align}
f_a(\vb r,\vb k)
&\simeq
\int_0^\infty d\tau\,e^{-\tau/\tau_J}
\mathcal K_a^{\Gamma(2)}(\vb r-\vb v\tau,\vb k),
\label{eq:transportKernel}\\
&\simeq
\tau_J\mathcal K_a^{\Gamma(2)}(\vb r,\vb k)
-\tau_J^2v_i\partial_i\mathcal K_a^{\Gamma(2)}(\vb r,\vb k)
+\cdots .
\label{eq:transportExpansion}
\end{align}
The corresponding transverse spin current is
\begin{align}
j_{s,z}^{a}
&=
\intk v_z f_a
\nonumber\\
&\simeq
\tau_J\intk v_z\mathcal K_a^{\Gamma(2)}
-\tau_J^2\intk v_zv_i\partial_i\mathcal K_a^{\Gamma(2)}
+\cdots .
\label{eq:end_current_gradient_expansion}
\end{align}
The first (\(\tau_J\)) term is the local \(W=v_z\) projection discussed in Eq.~\eqref{eq:finiteDirect}; it vanishes for a parabolic dispersion and clean continuum integration.
The second (\(\tau_J^2\)) term carries the nonlinear weight \(v_zv_i\); for the force-driven scalar source of Eq.~\eqref{eq:deltaFx}, however, it too vanishes by \(k_x\) parity in the clean parabolic limit. Integrating by parts twice, \(\partial_{k_x}\partial_{k_z}(v_zv_i)\) is constant for a parabolic band, leaving \(\intk\delta f_0^{(F_x)}\propto\intk v_x(-\partial_\varepsilon f_{\rm eq})\), which is odd in \(k_x\).
A finite momentum-integrated transverse spin current therefore appears only at higher order in \(\tau_J\), or through a nonparabolic dispersion, finite aperture, or boundary conversion that renders the weight nonlinear.
Before spatial integration, the spin texture set by \(\mathcal K_a^{\Gamma(2)}(\vb r,\vb k)\) stays finite without momentum integration.  A conventional whole-cloud time-of-flight image instead integrates this local contribution over \(\vb r\); consequently, the mixed derivative can cancel.  The clean-parabolic-gas readout is therefore a spatially windowed time-of-flight texture or an explicitly boundary-converted signal.
The local curvature kernel, the windowed time-of-flight texture, a propagated spin current, and a spin-density moment are thus related yet distinct objects.

The spin-density moment follows directly from the continuity equation.
For a closed cloud with vanishing boundary spin flux and a time-independent reference normalization \(N\),
\begin{align}
\partial_tD_z^a
&=
\frac{1}{N}\int d\vb r\,j_{s,z}^a
\nonumber\\
&{}
+
\frac{1}{N}
\int d\vb r\,z
\left(
\mathcal S_a^{\mathrm{loc}}
+\mathcal S_a^{\Lambda(1)}
+\mathcal S_a^{\Gamma(1)}
+\mathcal S_a^{\Gamma(2)}
\right).
\label{eq:end_dipole_evolution}
\end{align}
The current term is the bulk contribution generated when the divergence in Eq.~\eqref{eq:spinContinuity} is integrated by parts; the no-flux boundary condition removes only the surface term. If \(D_z^a\) is instead normalized by an instantaneous atom number \(N(t)\), the right-hand side of Eq.~\eqref{eq:end_dipole_evolution} acquires the additional term \(-[\partial_tN(t)/N(t)]D_z^a\).
In a clean parabolic gas, the bulk integral of the local dissipative curvature source may vanish for simple weights.
The measured \(D_z^a\) can nevertheless be generated by finite-time propagation, boundary conversion, finite apertures, or a nonparabolic optical-lattice dispersion.
This is why the main text treats \(s_a(\vb r,t)\), \(D_z^a\), \(j_{s,z}^a\), and spin-resolved momentum textures as separate readouts.

\paragraph{A Scenario for Gradient Spintronic Materials---}
The source/readout separation used above provides a controlled language for gradient-material spin conversion~\cite{Okano2019,Horaguchi2025}, for which a finite-film phenomenology can be written as follows.
A \(z\)-dependent spin-orbit-collision or spin-relaxation profile creates a spin-\(y\) source, while the finite device selects the measured signal through \(x\)-asymmetric boundaries~\cite{Okano2019,Horaguchi2025}.
The charge current flows along \(x\).
Oxidation, composition, mobility, and current-density profiles vary mainly along \(z\).
The bulk film is taken to be translationally invariant along \(x\).

Extrinsic spin-orbit scattering in the current-carrying state can generate a \(z\)-dependent spin source
\begin{align}
\mathcal C_y^{\nabla\mathrm{ext}}(\vb k,z)
&=
\mathcal A_y(z)
v_xv_z
\left[-\partial_\epsilon f_{\rm eq}(\epsilon_{\vb k})\right]
+\cdots ,
\label{eq:spintronics_bulk_source}
\end{align}
where \(\mathcal A_y(z)\) is the amplitude of the projected spin-\(y\) collision source.
For a current profile \(j_x(z)=\sigma(z)E_x\), a representative dependence is \(\mathcal A_y(z)\propto E_x\partial_z[\chi_{\mathrm{ext}}(z)\sigma(z)]\), with additional finite-film sensitivity to spin-relaxation gradients and asymmetric boundary absorption.

The direct measurement is described by a spin-\(y\) readout weight \(P_y(x,\vb k)\).
It is the weight with which a spin-\(y\) distribution at position \(x\) and momentum \(\vb k\) contributes to the measured \(z\)-directed spin signal:
\begin{align}
J_{s,z}^{y,\mathrm{meas}}
&=
\int dx\,dz\,\intk
v_z
P_y(x,\vb k)
\mathcal C_y^{\nabla\mathrm{ext}}(\vb k,z).
\label{eq:spintronics_readout_signal}
\end{align}
Here \(P_y\) is an effective boundary readout weight.
Left-right asymmetric contacts, spin sinks, or oxidized and metal-to-insulator ends give it a \(v_x\)-odd part through spin absorption, current crowding, extrinsic spin-orbit scattering, and, when heat removal is asymmetric, local heating:
\begin{align}
P_y(x,\vb k)
&=
P_y^{(0)}(x,\epsilon_{\vb k})
+\eta_y(x,\epsilon_{\vb k})v_x
+\cdots .
\label{eq:spintronics_readout_weight}
\end{align}
The \(v_x\)-odd term in Eq.~\eqref{eq:spintronics_readout_weight} selects the \(v_x\) factor in Eq.~\eqref{eq:spintronics_bulk_source}.
The measured direct signal then contains
\begin{align}
J_{s,z}^{y,\mathrm{meas}}
&\simeq
\int dx\,dz\,
\mathcal A_y(z)
\intk
\eta_y(x,\epsilon_{\vb k})
v_x^2v_z^2
\left[-\partial_\epsilon f_{\rm eq}(\epsilon_{\vb k})\right].
\label{eq:spintronics_projected_signal}
\end{align}

The inverse spin-injection measurement uses a scalar readout weight \(P_0(x,\vb k)\) for the charge-current signal:
\begin{align}
J_x^{\mathrm{inv,meas}}
&=
\int dx\,dz\,\intk
v_x
P_0(x,\vb k)
\mathcal C_0^{\nabla\mathrm{ext}}[f_y^{\mathrm{inj}}](\vb k,z).
\label{eq:spintronics_inverse_readout}
\end{align}
Here \(\mathcal C_0^{\nabla\mathrm{ext}}[f_y^{\mathrm{inj}}]\) is the scalar source generated by the injected spin-\(y\) distribution.
The direct signal is controlled by \(P_y\), whereas the inverse charge signal is controlled by \(P_0\).
This comparison concerns measured finite-device signals:
the two experiments project different combinations of source amplitudes, propagation lengths, spin relaxation, boundary absorption, and contact asymmetry, rather than a single bulk coefficient with one shared weight.
Material parameters that make the \(v_x\)-odd part of \(P_y\) strong while leaving the corresponding part of \(P_0\) weak, or vice versa, can produce the direct-inverse asymmetry measured in gradient-material films~\cite{Okano2019,Horaguchi2025}.

Recent solid-state quantum kinetic theory treats a related edge problem in weakly disordered multiband metals: boundary gradients of the longitudinal electron flow generate local electron angular-momentum density, including an orbital contribution connected to charge-current vorticity~\cite{ValetRaimondi2025PRB,Valet2025PRL}.
This edge angular-momentum mechanism is distinct from the cold-atom source derived here, where Moyal-product terms from coherent self-energy gradients and dissipative self-energy curvature generate momentum-space spin textures and, after the specified detector weight or propagation kernel is applied, spin-density or spin-current signals.
In gradient spintronic materials, both effects may enter a microscopic description of finite-film spin and angular-momentum signals.

\end{document}